\documentclass{ceab}   

\usepackage{epsfig}     
\usepackage{graphicx}   

\usepackage{ceabbib}     
\usepackage[T1]{fontenc}
\usepackage{textcomp}
\begin{document}
\countdef\pageno=0
\pageno=1

\def\tit{Solar rotation elements from sunspot observations by Ru{\dj}er Bo\v{s}kovi\'c}
\def\aut{M. HUSAK et al.}
\def\str{1--7}

\title{Solar rotation elements \lowercase{\textbf{\textit{i}}}, $\mathbf\Omega$ and period determined using sunspot observations by Ru{\dj}er Bo\v{s}kovi\'c in 1777\protect\footnote{\footnotesize C\lowercase{ontributed paper (oral presentation) at the} I\lowercase{nternational} A\lowercase{dvent} W\lowercase{orkshop 17th - 18th} D\lowercase{ecember 2020}, Z\lowercase{agreb} - G\lowercase{raz}, H\lowercase{var} O\lowercase{bservatory}, F\lowercase{aculty of} G\lowercase{eodesy}, U\lowercase{niversity of} Z\lowercase{agreb}, C\lowercase{roatia and} IGAM, I\lowercase{nstitute of} P\lowercase{hysics}, K\lowercase{arl}-F\lowercase{ranzens} U\lowercase{niversity} G\lowercase{raz}, A\lowercase{ustria}.}}

\author{M. HUSAK$^1$, R. BRAJ\v{S}A$^2$ and D. \v{S}POLJARI\'C$^2$\\
           \vspace{2mm}\\
\it$^1$Trako\v{s}\'canska 20, 42000 Vara\v{z}din, Croatia\\
\it$^2$Faculty of Geodesy, University of Zagreb, Ka\v{c}i\'{c}eva 26, \\
\it 10000 Zagreb, Croatia
}

\maketitle


\begin{abstract}
Ru{\dj}er Bo\v{s}kovi\'c developed methods for determination of solar rotation elements: the solar equator inclination $i$, the longitude of the node $\Omega$ and the period of solar rotation. In his last work {\it Opera pertinentia ad opticam et astronomiam}, published in 1785, in the chapter {\it Opuscule II} he described his methods, the formulae with figure descriptions and an example for calculation of the solar rotation elements with detailed numerical explanation using his own observations performed in September 1777. The original numerical procedure was performed using logarithmic formulae. In present work we give a description of the original results of R. Bo\v{s}kovi\'c and compare them with our recalculated values.
\end{abstract}


\section{Introduction}

Ru{\dj}er Bo\v{s}kovi\'c (1711--1787) was a Croatian Jesuit priest and his main fields of interest were physics, mathematics, geodesy, cartography, and astronomy, including both theoretical and practical aspects. Bo\v{s}kovi\'c performed research on determination of the solar rotation elements which are called Carrington's elements: the longitude of the ascending node $\Omega$ and the inclination of the solar equator to the ecliptic $i$ (Figure 1).

\begin{figure}
\begin{center}
\epsfig{file=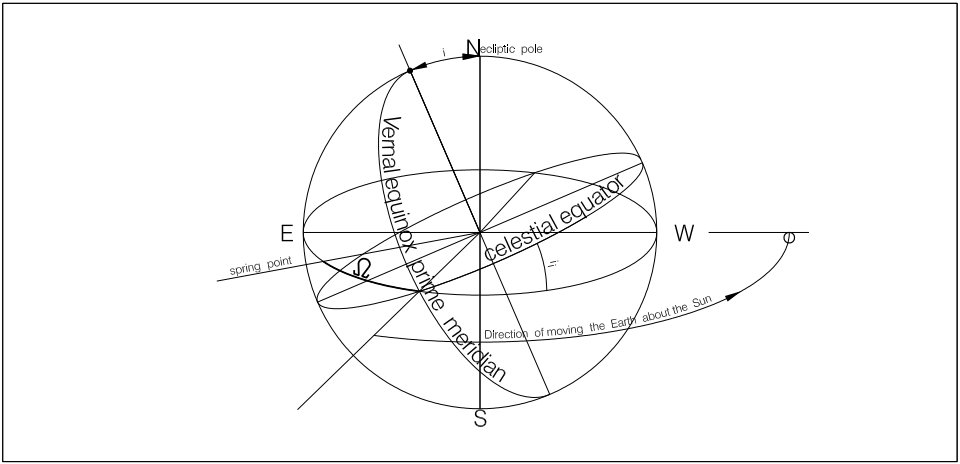,width=\textwidth}
\end{center}
\caption{The Carrington solar rotation elements: the longitude of the ascending node, $\Omega$, and the inclination of the solar equator regarding ecliptic, $i$ (according to Figure 4-2 in Braj\v{s}a (1989)).
   }
\label{fig1}
\end{figure} 

In his first dissertation {\it De maculis solaribus} (Boscovich, 1736) he described his methods for solar rotation elements determination using sunspot observations. In September 1777 he observed sunspots in Noslon near Sens 120 km south from Paris where he was a guest of the cardinal Paul d'Albert de Luynes (1703--1788), an amateur astronomer. The detailed research report about the solar observations was published in {\it Opuscule II} in fourteen chapters and appendix {\it Appendice} (Boscovich, 1785, 75-178). The {\it Opuscule II} gives (in the chapters {\it\S. I.} -- {\it\S. XIV.}) as follows: the formulae with figure descriptions ({\it Pr\'eface}) and observations ({\it\S. I.}), the sunspot positions determination 
({\it\S. II.} -- {\it\S. III.}), the methods ({\it\S. IV.} -- {\it\S. VII.}), the calculation instructions ({\it\S. IX.} -- {\it\S. XIII.}), and additionally the reflexions ({\it R\'eflexions} in {\it\S. VIII.}  and {\it\S. XIV.}).

\section{Methods}

In September 1777 Bo\v{s}kovi\'c observed four sunspots using the telescope\footnote{T\'elescope dans Lalande T2 - Noslon - Wikip\'edia (wikipedia.org). https://fr.wikipedia.org/wiki/Noslon\#/media/Fichier:T\'elescope\_dans\_Lalande\_T2.png, accessed 11. 05. 2021.} (Figure 2) with optical micrometre and pendulum for time measurement. As an example, he used observations of one sunspot. Cardinal de Luynes had contemporary and professional instruments, which his guests used for research.

\begin{figure}[ht]
\begin{center}
\epsfig{file=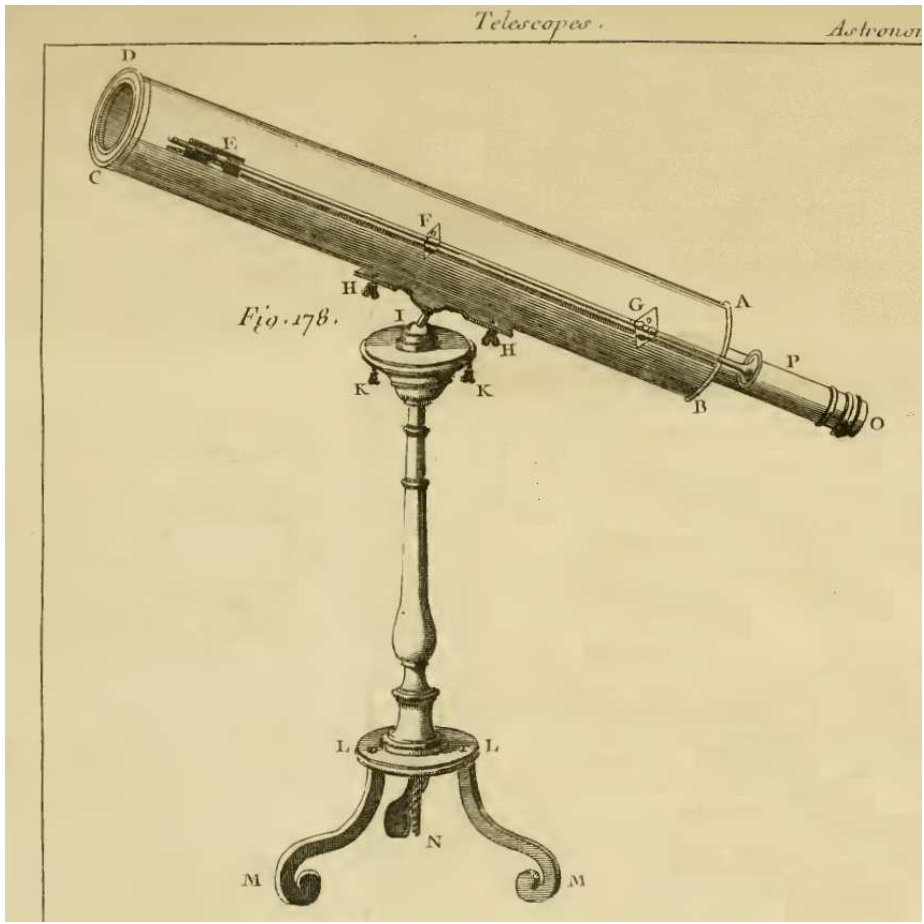,width=12cm}
\end{center}
\caption{Telescope that was described by Lalande in {\it Astronomie T II.}}
\label{fig2}
\end{figure} 

For his methods Bo\v{s}kovi\'c developed step-by-step procedures in a form of twelve tables which are convenient for calculations with logarithmic tables (Boscovich, 1785, {\it Tab. I.} to {\it Tab. XII.}, 166-169). He performed measurements and data reduction in two steps using his own formulae (Boscovich, 1785, {\it Pr\'eface}, 	\textnumero 12 -- \textnumero 18, 81-84). First, he measured the sunspot position at the solar disk and the true solar time, as well as the position of the sunspot in ecliptic coordinate system, the longitude and the latitude ({\it T.M.}, {\it lon.t}, {\it lat.B.t}) (Boscovich, 1785, {\it Tab. I.} and {\it Tab. II.}, 166-167), in the moment of the sunspot observation (Boscovich, 1785, {\it\S. I.}, 87-89). Then he calculated the solar rotation elements: $i$, $\Omega$ and the sidereal and the synodic rotational period. 
The Bo\v{s}kovi\'c's methods use the mean solar time and sunspot position combinations ({\it T.M.}, {\it lon.t}, {\it lat.B.t}) for solar rotation elements determination: the longitude of the node $N=\Omega$ using two positions of the same sunspot in {\it Tab. III.} and {\it Tab. IV.} and the solar equator inclination $i$ using two positions of the same sunspot and the longitude of the node $N=\Omega$ in {\it Tab. V.} and {\it Tab. VI.} The periods of the solar rotation, the sidereal and synodic ones, are given in {\it Tab. VII.} to {\it Tab. XI.} The determination of the longitude of the node $N=\Omega$ and the solar equator inclination $i$ using three positions of the same sunspot are presented in {\it Tab. XII.} These tables are reproduced with detailed descriptions in the paper by Husak, Braj\v{s}a, and \v{S}poljari\'c (2021).

\section{Results}

Bo\v{s}kovi\'c used observational data, astronomic almanac data and logarithmic tables for the determination of the mean solar time and positions of a sunspot. Observational data are given in {\it Opuscule II} and other input data for calculation were taken from astronomic almanac. The values from the astronomical almanac {\it Connoissance des temps}\footnote{URL 1: Connoissance des temps. Connoissance des temps, Pour l' Ann\'ee commune 1777. publi\'ee Par l'ordre de l' Acad\'emie Royale des Sciences, el calcul\'ee Par M. Jeaurat, de la m\^eme Acad\'emie. A Paris de l'imprimerie royale. M. DCCLXXVI. (Observatoire de Paris), https://gallica.bnf.fr/ark:/12148/cb327469896/date} are: the correction of the local meridian to Paris meridian $\Delta t_{\rm Sens}=3^m48^s$ at the page 268, the correction of the true solar time to the mean solar time $\Delta t_{\rm MST}$ at the page 103 ({\it T.M., French Temps moyen au Midi vrai} -- Average time at true noon), the longitude 
$\lambda_\odot$=$lon._\odot$ and the declination $\delta_\odot$=dec.$_\odot$ of the Sun in Paris interpolated to the true solar time in minutes for a day of observation at the pages 102-103, the apparent diameter of the Sun $D_\odot$ interpolated to the day of observation at the page 108 and the inclination of the ecliptic rounded to whole minutes $\varepsilon=23^\circ 28'$ at the page 4. Ru{\dj}er Bo\v{s}kovi\'c determined constant of the micrometre C=1237/1915 (relative units of micrometre per arc sec) experimentally by multiple observations of the vertical size of the solar disc on 11th September 1777 using interpolated solar diameter $D_\odot$. 
In present work we recalculated results using original Bo\v{s}kovi\'c's and corrected mean solar time and positions ({\it T.M.}, {\it lon.t}, {\it lat.B.t}). In that way we can compare the results of the example of Bo\v{s}kovi\'c and the present work results. The input data for the solar rotation elements are data for the sunspot ({\it T.M., lon.t, lat.B.t}) and, depending on the method, the solar rotation elements determined earlier ($\Omega=N$, $i$, $T$).
Bo\v{s}kovi\'c determined the longitude of the node $\Omega=N$ and the solar equator inclination $i$ combining different pairs of observational data. So, he obtained the following values: as the arithmetic mean of six pairs $\Omega_6=70^\circ 21'$, of eight pairs $\Omega_8=71^\circ 32'$ and of ten pairs $\Omega_{10}=73^\circ 09'$ of the positions of the same sunspot which Bo\v{s}kovi\'c used in the {\it Tab. IV.} Further, the solar equator inclination $i=7^\circ 44'$ was calculated as the arithmetic mean of five pairs of the positions of the same sunspot which Bo\v{s}kovi\'c used in the {\it Tab. VI.} The sidereal period $T'=26.77$ days was determined as the arithmetic mean of six pairs of times and sunspot positions of the same sunspot which Bo\v{s}kovi\'c used in the {\it Tab. X.} and synodic period $T''=28.89$ days in the {\it Tab. XI.} The longitude of the node $N=\Omega_{136}=74^\circ 03'$ and the solar equator inclination $i=6^\circ 49'$ using three positions (1, 3, 6) of the same sunspot are given in the {\it Tab. XII.}
We recalculated the results using the methodology of Bo\v{s}kovi\'c as follows. The longitude of the node $\Omega=N$ was calculated as the arithmetic mean of six pairs $\Omega_6=70^\circ 21'$, eight pairs 
$\Omega_8=71^\circ 32'$ and ten pairs $\Omega_{10}=73^\circ 12'$ of the positions of the same sunspot which Bo\v{s}kovi\'c used in the {\it Tab. IV.} The solar equator inclination $i=7^\circ 45'$ was calculated as the arithmetic mean of five pairs of the positions of the same sunspot which Bo\v{s}kovi\'c used in the {\it Tab. VI.} The sidereal period $T'=26.76$ days was calculated as the arithmetic mean of six pairs of corrected times and sunspot positions of the same sunspot which Bo\v{s}kovi\'c used in the {\it Tab. X.}
The corresponding synodic period is $T''=28.87$ days and with Bo\v{s}kovi\'c's times $T'=26.77$ days and $T''=28.89$ days. The longitude of the node $N=\Omega_{136}=74^\circ 03'$ and the solar equator inclination $i=6^\circ 48'$ using three positions (1, 3, 6) of the same sunspot are given in the {\it Tab. XII.}

\section{Analysis and discussion}

In modern research we do not use logarithmic tables and step-by-step calculations which Bo\v{s}kovi\'c used in his example (Boscovich, 1785, {\it Tab. I.} to {\it Tab. XII.}, 166-169). Modern computers enable fast complex calculations in double precision. This research is performed in spreadsheet 
(Microsoft Excel\textsuperscript\textregistered). It could be performed at any other computer platform which supports elemental mathematical functions.
The main problem in reproduction of Bo\v{s}kovi\'c's example was discovering calculation chains at the places where formulae are missing. The formulae were successfully reproduced using Bo\v{s}kovi\'c's results and descriptions of calculations in {\it Opuscule II.}
We determined the mean solar time {\it T.M.} using correction for the mean solar time which resulted with {\it T.M.} values different from these which Bo\v{s}kovi\'c published. We assume that Bo\v{s}kovi\'c used a wrong table from the astronomical almanac, but we do not have the final confirmation for this hypothesis yet.

\section{Conclusion}

In present work we successfully reproduced the sunspot positions and the solar rotation elements using the original formulae and original observational data of R. Bo\v{s}kovi\'c. The reproduced values of the mean solar time and the positions ({\it T.M.}, {\it lon.t}, {\it lat.B.t}) are almost the same. We then repeated determination of the solar rotation elements with Bo\v{s}kovi\'c's original coordinates and time and additionally periods with corrected time.
The reproduction of the results uses the methodology which Bo\v{s}kovi\'c used in his example with the same input data for the time and sunspot positions. The present work results confirm that Bo\v{s}kovi\'c did extraordinary work, from the beginning (his own methods, with his own observations and calculations) to the final results. The example confirms and verifies his methods for solar rotation elements using his own sunspot observations performed in the late 18th century with very similar results compared with present work results.
The most important achievement of the present work is a confirmation of the methodology and reproduction of the results with original formulae for the methods for the solar rotation elements determination using sunspot observations of Ru{\dj}er Bo\v{s}kovi\'c. In September 1777 Bo\v{s}kovi\'c observed four sunspots which we can now further reduce and apply modern formulae convenient for performing the methods of Ru{\dj}er Bo\v{s}kovi\'c. 
This research is in progress and in the next step we will analyse the precision of input parameters (solar ephemeris, apparent solar diameter, etc.) on the precision of output data in solar rotation determination. Further, we will develop formulae applicable for modern computer programs and try to fix numerical calculation problems and difficulties. These results will enable application of the methods for solar rotation determination to modern data and will give a possibility to compare the results of R. Bo\v{s}kovi\'c with modern results.

\section*{Acknowledgements}

This work has been supported by the Croatian Science Foundation under the project 7549 "Millimeter and submillimeter observations of the solar chromosphere with ALMA".

\section*{References}

\begin{itemize}
\small
\itemsep -2pt
\itemindent -20pt
\item [] Boscovich, R. J.: 1736, {\it A. M. G. D. De M\lowercase{ACULIS} S\lowercase{OLARIBUSN} 
\lowercase{EXERCITATIO ASTRONOMICA HABITA} (About sunspots), Colegio Romano Societatis Jesu}, Typographia Komarek, Romae, 3-10, (in Latin).

\item [] Boscovich, R. J.: 1785, {\it Opera pertinentia as opticam, et astronomiam (... optics and astronomy), Maxima ex parte nova, \& omnia hucusque inedita, In quinque Tomos distributa Ludovico XVI. Galliarum Regi Potentissimo Dicata.} Tomus quintus, {\it Opuscule II}, Bassani, 75-178. (in French).

\item [] Braj\v{s}a, R.: 1989, {\it Determination of the solar differential rotation at high latitudes using polar crown filaments, Diploma work}, University of Zagreb, Faculty of Science and Mathematics, Zagreb, (in Croatian).

\item [] Husak, M., Braj\v{s}a, R., \v{S}poljari\'c, D.: 2021, {\it RGN Zbornik} {\bf 36} No. 3, in press.
\end{itemize}


\bibliographystyle{ceab}
\bibliography{Brajsa_ceab_2020}

\end{document}